\documentclass[%
 prA,showkeys,superscriptaddress,
 amsmath,amssymb,
 reprint,%
]{revtex4-1}

\usepackage{graphicx}
\usepackage{dcolumn}
\usepackage{bm}

\usepackage[utf8]{inputenc}
\usepackage[T1]{fontenc}
\usepackage{mathptmx}

\usepackage{color}
\usepackage{subfigure}
\usepackage{paralist}
\usepackage{appendix}
\usepackage{amsmath}
\usepackage {graphicx}
\graphicspath{{../../figs/}}

\newcommand {\qfig}[1]{Fig.~\ref{#1}}

\newcommand {\queq}[1]{(\ref{#1})}
\newcommand {\qeq}[1]{Eq.~\queq{#1}}

\newcommand {\beql}[1]{\begin{equation} \label{#1}}
\newcommand {\eeql}{\end{equation}}
\newcommand {\beq}{\begin{equation}}
\newcommand {\eeq}{\end{equation}}

\newcommand {\etal}{\emph{et al.}\ }
\newcommand {\Rpl}{R_{\rm pl}}

\begin{document}

\title{Structure and size of the plastic zone formed during nanoindentation of a metallic glass}

\author{Karina E. Avila}
\affiliation{%
Physics Department,
Technical University,
Erwin-Schr{\"o}dinger-Stra{\ss}e, D-67663 Kaiserslautern, Germany}
\affiliation{%
Research Center OPTIMAS,
Technical University,
Erwin-Schr{\"o}dinger-Stra{\ss}e, D-67663 Kaiserslautern, Germany}

\author{Stefan K\"uchemann}
\affiliation{%
Physics Department,
Technical University,
Erwin-Schr{\"o}dinger-Stra{\ss}e, D-67663 Kaiserslautern, Germany}

\author{Herbert M.~Urbassek}
\email{urbassek@rhrk.uni-kl.de}
\homepage{http://www.physik.uni-kl.de/urbassek/}
\affiliation{%
Physics Department,
Technical University,
Erwin-Schr{\"o}dinger-Stra{\ss}e, D-67663 Kaiserslautern, Germany}
\affiliation{%
Research Center OPTIMAS,
Technical University,
Erwin-Schr{\"o}dinger-Stra{\ss}e, D-67663 Kaiserslautern, Germany}

\date{\today}

\begin{abstract}

Using molecular dynamics simulation, we study the plastic zone created during nanoindentation of a large CuZr glass system. 
The plastic zone consists of a core region, in which virtually every atom undergoes plastic rearrangement, and a tail, where the density distribution of the plastically active atoms decays to zero. 
Compared to crystalline substrates,  the plastic zone in metallic glasses is significantly smaller than in crystals. The so-called plastic-zone size factor, which relates the radius of the plastic zone to the contact radius of the indenter with the substrate, assumes values around 1, while in crystals -- depending on the crystal structure -- values of 2--3 are common. The small plastic zone in metallic glasses  is caused   by the essentially homogeneous deformation in the amorphous matrix, while in crystals heterogeneous dislocations prevail, whose growth leads to a marked extension of the plastic zone.

\end{abstract}


\maketitle

Compared to crystals, glasses show a much larger elastic elongation which is around 2\% \cite{HSF16}; this feature   makes them promising materials for engineering applications. The understanding of the underlying mechanisms for enhancing the elastic limit and the plastic ductility of metallic glasses is  an active area of research \cite{HSF16,HSW*08,MD18}. In principle, the deformation mechanism of glasses    exhibits different features as compared to crystals. 
In the latter, deformation occurs via the movement of defects in the crystalline structure called dislocations \cite{Nab67,CN16}. In the former, the deformation occurs via cooperative atomic rearrangements which cause strain localization called shear bands \cite{ML15,KLD*18,KPA*14}. 

In crystals, the damage produced inside the sample due to nanoindentation is assumed to be hemispherical with a radius $\Rpl$ \cite{Joh85book,DBG05,ARU18}. This radius scales with the contact area of the indenter $a_c$ as 
\begin{equation}
\Rpl=fa_c ,
\label{pl_factor}
\end{equation}
where $f$ is called the plastic-zone size factor. The size factor $f$ is used to characterize the spatial extension of the damage introduced into the material \cite{DBG05,ARU18}. Recently, molecular dynamics (MD)   simulations performed for a variety of crystals found $f$ to be in the range of 2.5--3.5 \cite{ARU18}. For metallic glasses, however, $f$ and $\Rpl$ have apparently not been discussed up to now. 

The purpose of this Letter is to study the damage created by nanoindentation using MD simulation of a CuZr-based  metallic glass. In order to do this, we identify the rearranging atoms and measure the changes in their  density  around the indenter. This allows us to determine  the size of the plastic zone in its dependence of temperature and  indentation depth. 


We use the open-source code LAMMPS \cite{Pli95} to simulate the binary-composition metallic glass Cu$_{64.5}$Zr$_{35.5}$. The sample consists of $N=5,619,712$ atoms contained in a cubic simulation box of length $L$. Its atomic density $\rho_0=N/L^3$ varies from 0.0616 for the lowest temperature (0.1 K) to 0.0599, for the highest temperature (1000 K) investigated here. The atomic interactions are modeled by the embedded-atom-method  potential developed by Mendelev \etal \cite{MSK07}. A crystalline mixture was first heated to a temperature above the melting point, $T=2000$ K, for 500 ps and then cooled to the final temperature with a quenching rate of 0.01~K/ps to obtain the metallic glass.  
The glass transition temperature for this particular composition and potential is slightly below 1000~K \cite{LJP12}. Here, we simulated samples at 8 different temperatures including a temperature above the glass transition, $T=1000$~K.

During the preparation of the sample, periodic boundary conditions were applied in all directions and an isobaric ensemble (at vanishing pressure) with a Nose-Hoover thermostat was used.
 Once the final temperature is reached, the system is left to relax for 200 ps with periodic boundary conditions; finally, the top surface is switched to free boundary conditions and the system is relaxed for another  300 ps. 10 atomic layers at the bottom of the sample are kept fixed in order to mimic the immobile bulk of the metallic glass in a real experiment and to avoid center-of-mass translation of the entire sample during indentation.

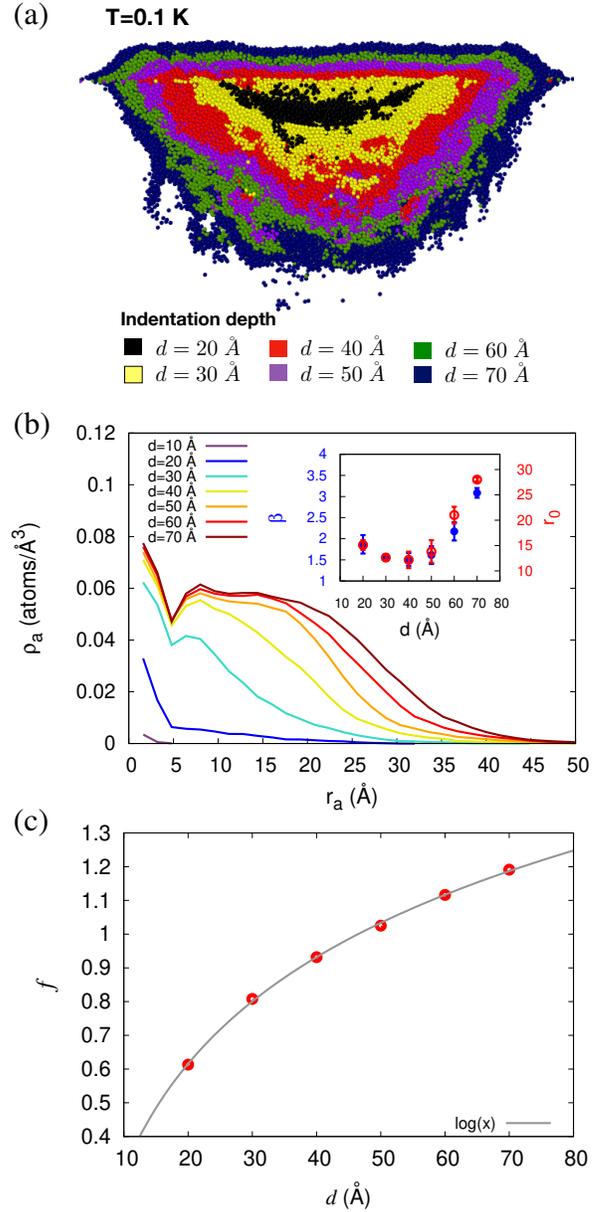
\begin{figure}
  \begin{center}
   \begin{picture}(220,450)
   \put(25,300){\includegraphics[width=0.37\textwidth]{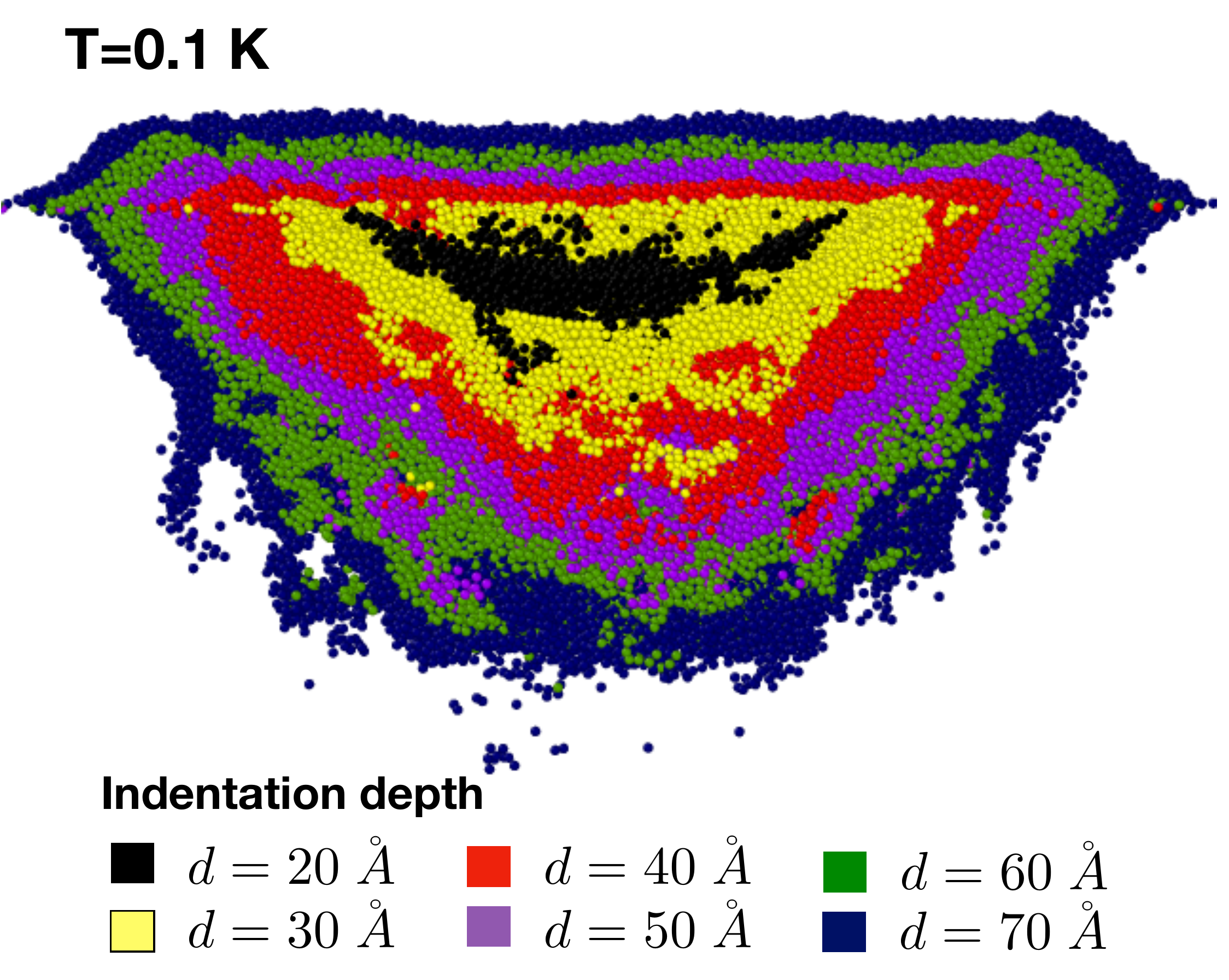}}
 \put(-2,133) {\includegraphics[width=0.45\textwidth]{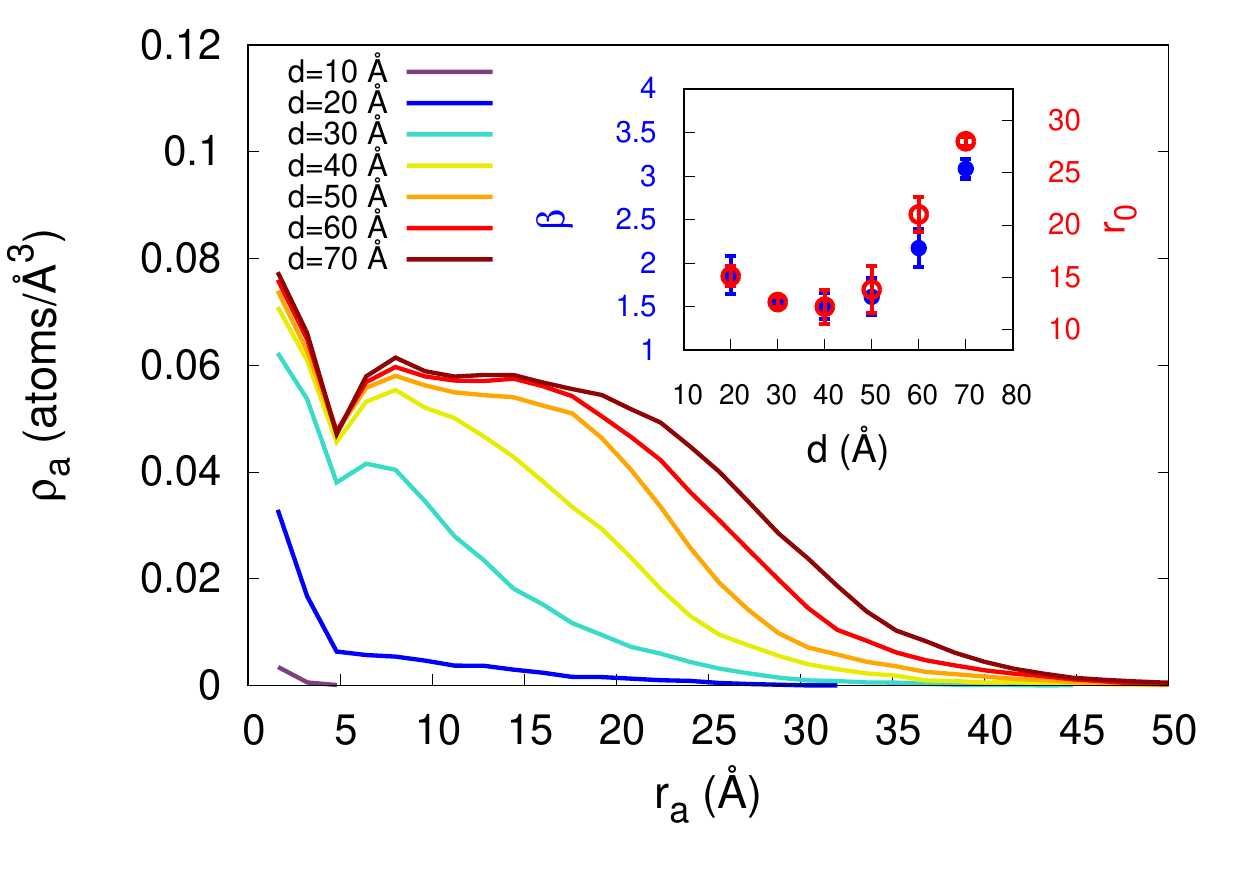}}
  \put(2,-15) {\includegraphics[width=0.44\textwidth]{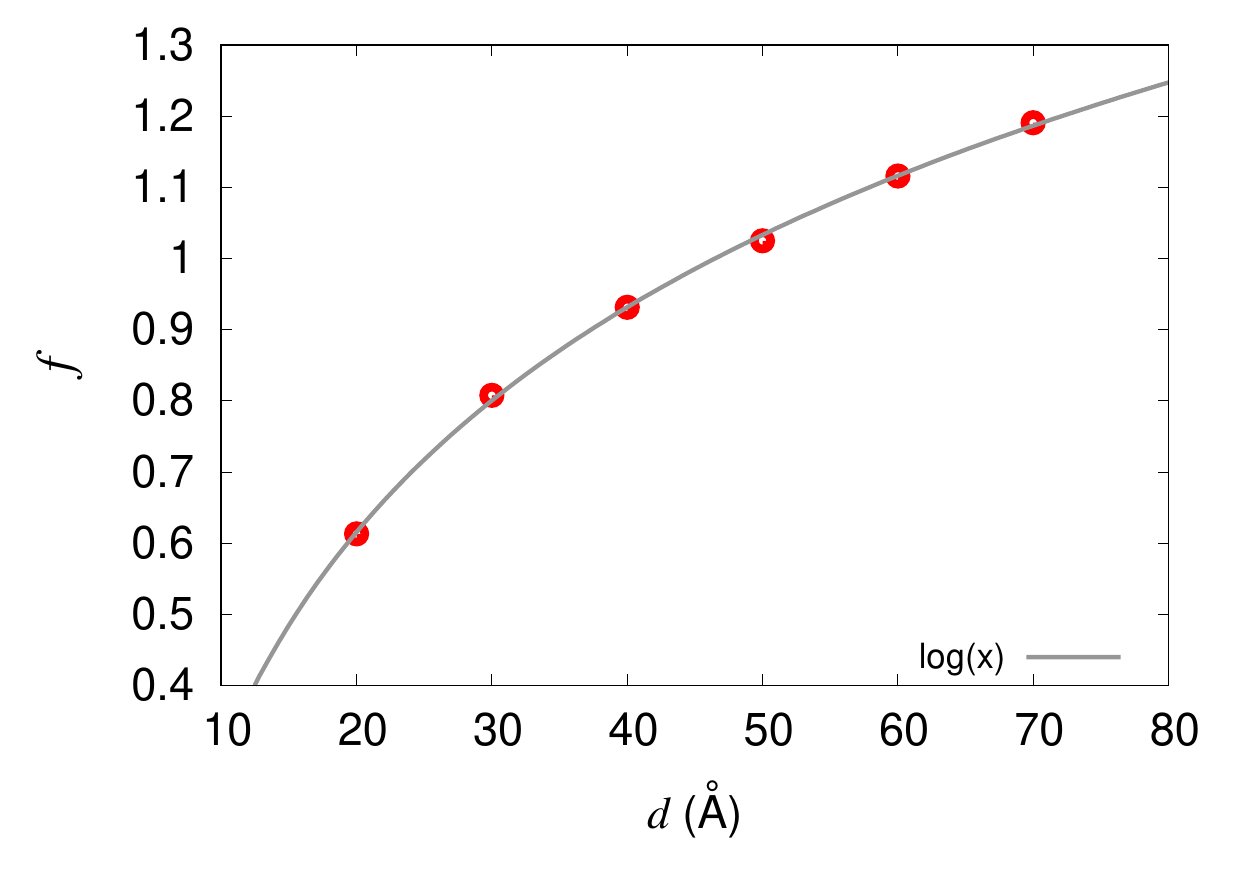}}
\put(0,440){\large{(a)}}
  \put(0,285){\large{(b)}}
  \put(0,135){\large{(c)}}
   \end{picture}
  \end{center}
\caption{Influence of indentation depth $d$ on plastic zone at a temperature of $T=0.1$ K. 
(a) Superimposed plastic zones for various indentation depths $d$. (b) Density of plastically active atoms as a function of the distance from the indenter tip, $r_a$, for various indentation depths. (c) Plastic-zone size factor $f$ as a function of indentation depth. The solid line shows a logarithmic increase. }
  \label{fig:Density-depth}
\end{figure}

The tip has a spherical form with a radius of $R=10$ nm. The purely repulsive force exerted by the tip on the system is given by $F(r)=K(r-R)^2$, ($r<R$), 
where $r$ is the distance of an atom to the center of the indenter. 
The stiffness constant of the tip has been set to $K=10$ eV/\AA$^3$. During indentation we keep the temperature fixed by using an NVT ensemble. 
In our simulation, first the tip is indented to its final depth $d$ at 20 m/s; finally, the indenter is removed with the same velocity. In all this study -- except \qfig{fig:Density-depth} and its discussion -- we use $d=7$~nm. 

The open source visualization tool OVITO \cite{Stu10} is used in our work to analyze and visualize the atomistic configurations.


\begin{figure}
  \begin{center}
{\includegraphics[width=0.38\textwidth]{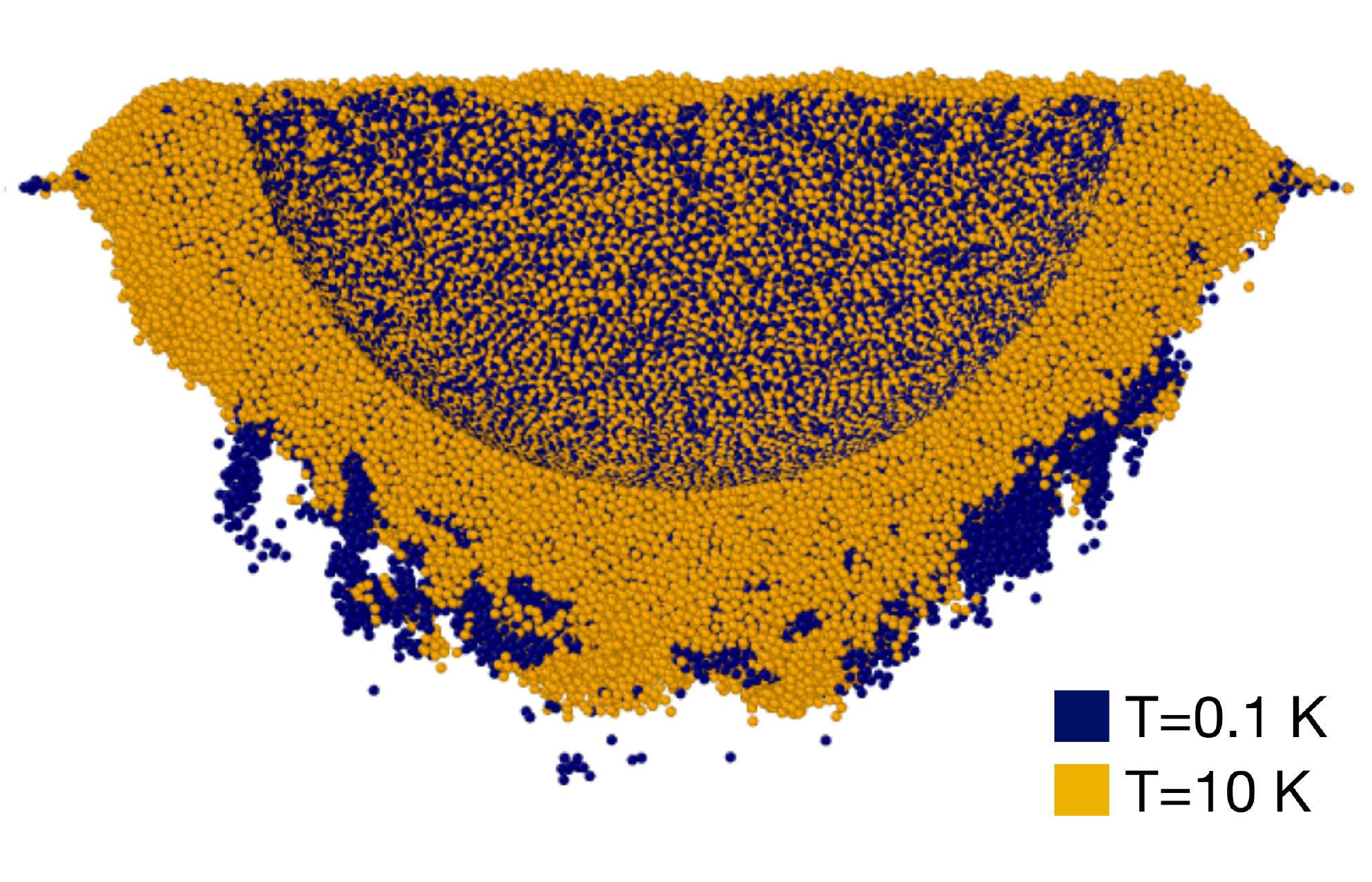}}
  \end{center}
\caption{Plastic zones for temperatures $T=0.1$ K and $T=10$ K at $d=70$ \AA; the zones are graphically superimposed in order to allow for easy comparison. This figure only shows atoms with VMSS$>0.3$. Indenter has been removed for clarity. Note the shear bands formed at $T=0.1$ K, but not at 10 K.}
  \label{fig:Plastic}
\end{figure}

In contrast to crystals \cite{ARU18}, the plastic zone produced by nanoindentation in  metallic glasses is quite isotropic around the indenter (see Fig.~\ref{fig:Plastic}). The definition of plasticity in metallic glasses is not as straightforward as for crystals, where  dislocations mark the  plastic deformation. In general, in amorphous materials the local plasticity may be accounted for by  the von-Mises shear strain (VMSS) \cite{SOL07} or by the non-affine squared displacement \cite{FL98}. Here, we use the VMSS. We shall denote atoms that exceed a certain threshold value VMSS$_{\text{cut}}$ as `plastically active atoms'. 
Fig.~\ref{fig:Plastic} shows a cross-sectional view of the plastic zone for a value of VMSS$_{\text{cut}}=0.3$. Data for temperatures $T=0.1$ K and $T=10$ K are superimposed to identify differences. For 0.1 K,  the presence of more inhomogeneous deformation, i.e., shear bands,  is obvious if compared to the closest simulated temperature, $T=10$ K. 

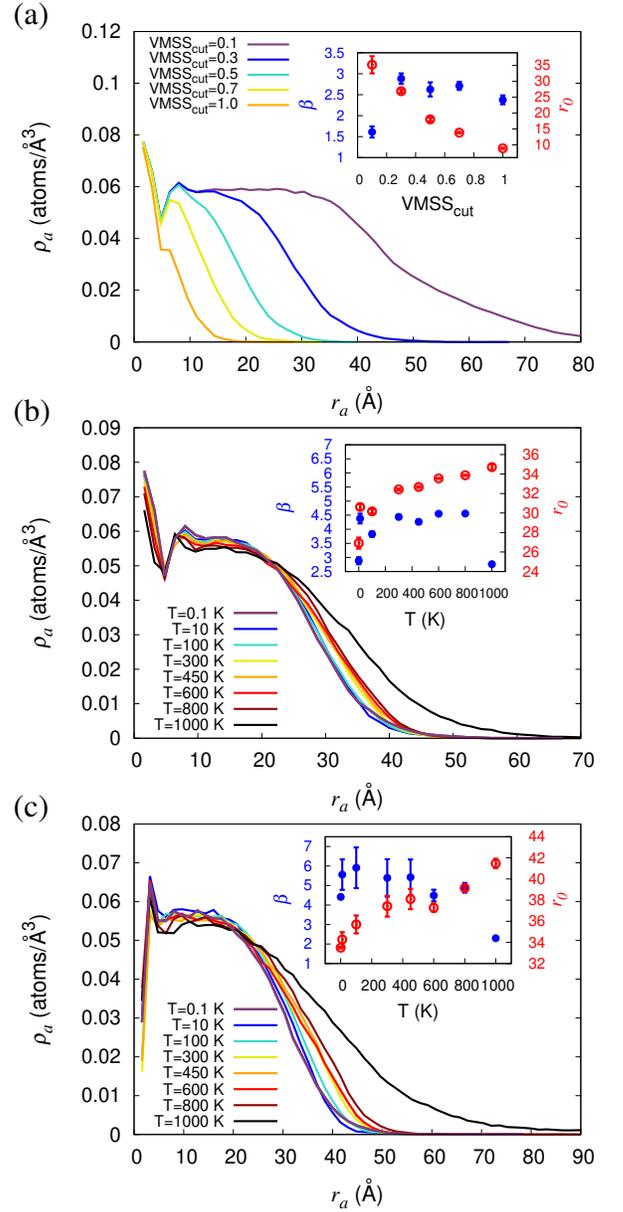
\begin{figure}
  \begin{center}
     \begin{picture}(220,450)
   \put(0,285) {\includegraphics[width=0.45\textwidth]{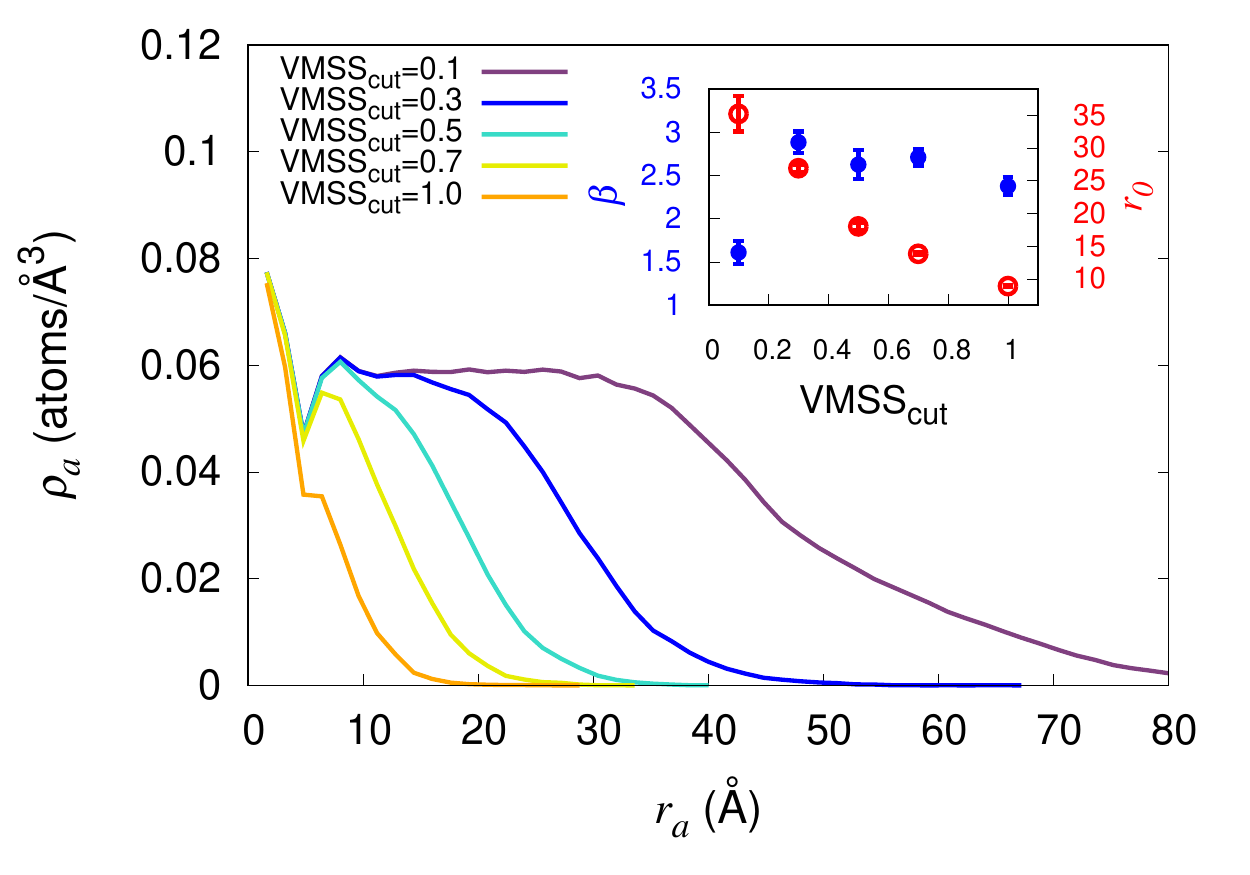}}
 \put(0,135){\includegraphics[width=0.45\textwidth]{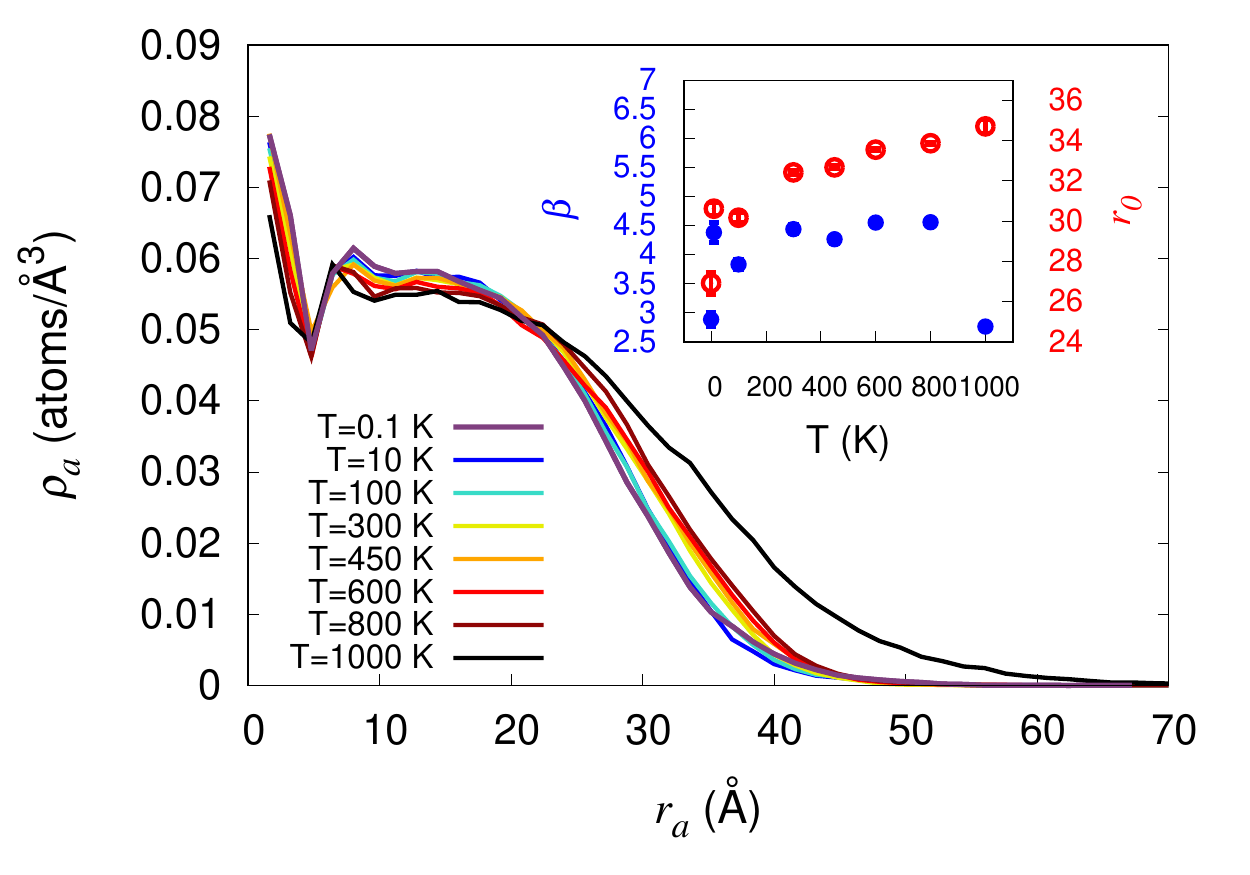}} 
  \put(0,-15) {\includegraphics[width=0.45\textwidth]{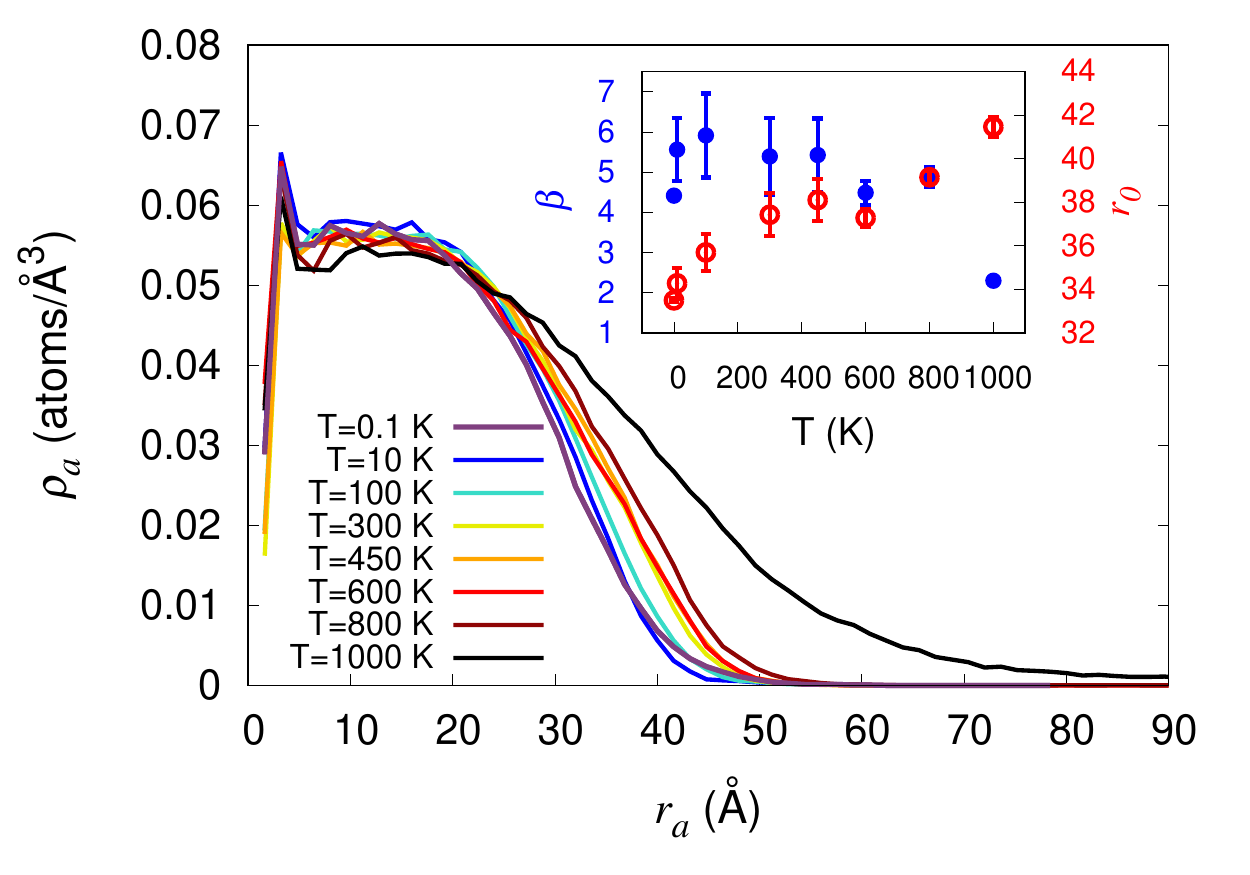}}
  \put(0,440){\large{(a)}}
  \put(0,290){\large{(b)}}
  \put(0,140){\large{(c)}}
   \end{picture}
  \end{center}
\caption{Density of the plastically active atoms, $\rho_a$, as a function of the distance from the indenter tip, $r_a$, for (a) various cut-offs of the von-Mises shear strain, VMSS$_{\text{cut}}$,  (b) various temperatures at full indentation, $d=70$ \AA, and (c)  after the indenter was removed. The insets in these figures show the  parameters $\beta$ and $r_0$ obtained from fitting to Eq.~(\ref{comp_ex}).}
  \label{fig:Density}
\end{figure}

We determine the density of the plastically active atoms, $\rho_a$ -- in short denoted as `plastic density' -- as a function of distance from the indenter surface, $r_a$. To this purpose,  the plastic zone (see Fig.~\ref{fig:Plastic}) is divided into spherical shells of 1.6~\AA\ thickness, and the number of plastically active atoms within the shell is divided by the volume of the shell. We exclude the atoms on the sample surface and the pileup formation from this calculation. Fig.~\ref{fig:Density} shows the plastic density $\rho_a$ as a function of the distance from the indenter surface $r_a$. 

We first discuss the influence of the  cut-off of the von-Mises shear strain, VMSS$_{\text{cut}}$, on the density distribution in Fig.~\ref{fig:Density}a (temperature of $T=0.1$ K). In all cases, immediately at the indenter surface, $r_a<3$ \AA, we observe a  shell with densities exceeding the atomic density of the glass, $\rho_0$; these are compressed atoms which had no opportunity to relax. Further outward, at around $r_a=5$ \AA, a minimum is observed which is caused by a packing mismatch between the compressed zone and the more relaxed plastic particles further outside. 
Beyond  this minimum, we can observe a plateau which signals the formation of a dense core of plastically active atoms close to the indenter. This densification is most noticeable for VMSS$_{\text{cut}}=0.1$ and is disappearing with increasing VMSS$_{\text{cut}}$, which is expected since less particles are considered in the calculation. The decay of the plastic density is very similar for all values of VMSS$_{\text{cut}}$, and can be fitted by a compressed exponential function,
\begin{equation}
\rho_a(r_a)=a \exp{\left[-\left(\frac{r_a}{r_0}\right)^{\beta}\right]} .
\label{comp_ex}
\end{equation}
For the fits, we  focus on the decaying tail of the distribution and fit for $r_a>15$~\AA.
The values of $\beta$ are around 2--2.5 indicating a Gauss-like decay, with the exception of VMSS$_{\text{cut}}=0.1$,  
where the value of $\beta$ is noticeably smaller [see inset of Fig.~\ref{fig:Density}(a)]. 
For the rest of the analysis performed in this manuscript, we use VMSS$_{\text{cut}}=0.3$. We select this value because the dense core is still noticeable for this choice and also because its decay follows the same behavior as for larger VMSS$_{\text{cut}}$. 

The temperature dependence of the plastic density is displayed  in Fig.~\ref{fig:Density}(b). The density decay is very similar for all temperatures with $\beta$ around 4.5, shown in the inset of Fig.~\ref{fig:Density}(b). Only for  the lowest temperature $T=0.1$ K discussed above and  for the temperature above the glass transition, $T=1000$ K, the decay is slower.  For the temperature above the glass transition, the slower decay of the plastic density can be explained by the admixture of thermal processes to the  mechanical relaxation occurring in the plastic zone.  This mixture facilitates the plastic rearrangement of the atoms deeper into the sample. However, the slower decay of the plastic density at $T=0.1$ K is not due to thermal activation but due to shear-band initiation. As discussed above  in Fig.~\ref{fig:Plastic},  the presence of  shear bands  is obvious at $T=0.1$ K  if compared to the slightly higher temperature of $T=10$ K. The formation of shear bands implies a higher degree of cooperativity among the atoms which might be attributed to the high plastic density at $T=0.1$ K. This interpretation is in agreement with the strongly heterogeneous deformation behavior during nanoindentation of dense systems such as crystals.

\begin{figure}
  \begin{center}
  {\includegraphics[width=0.45\textwidth]{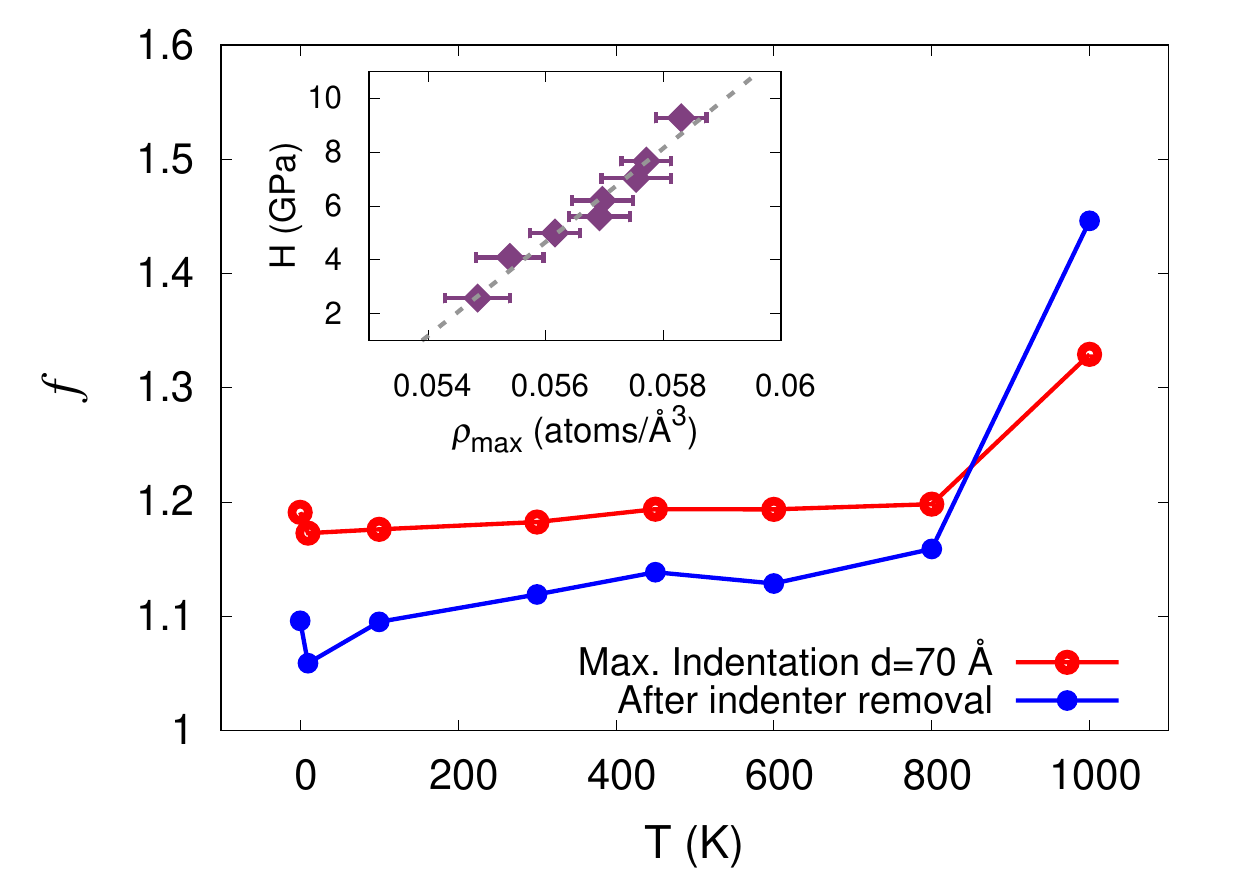}}
  \end{center}
\caption{Plastic-zone size factor $f$ as a function of temperature. {\em Inset:} Hardness \cite{AKAU19} as a function of the core plastic density, $\rho_{\text{max}}$.}
  \label{fig:f_factor}
\end{figure}

After the indenter is removed, the density of plastically active atoms will have changed and 
 a similar analysis must be performed. Since a temperature-dependent elastic and viscoelastic re-bound takes place \cite{AKAU19,AKAU19_scratch} the indent pit changes its form. However, we find here that it is still described by a spherical cap with a curvature radius of 10 nm, such that the same method as outlined above can be used to calculate the plastic density. 
 The results of the plastic-density calculations are shown in Fig.~\ref{fig:Density}(c). It is noticeable that the plastic-density distribution after  indenter removal decays considerably  faster than when the indenter is present,  as indicated by the enhanced values of the stretching exponent $\beta$. As the indenter was removed, also the stress field exerted by it has vanished and strained atoms return closer to their original positions; this explains the systematically smaller extension of the plastic density and the stronger decay after indenter removal, Fig.~\ref{fig:Density}(c), than before removal, Fig.~\ref{fig:Density}(b).  However, above the glass transition, the plastic-density decay is slower and therefore extends farther inside the sample; this is again due to the thermal activation occurring at this high temperature.

The extension of the plastic zone is determined from the maximum value, $r_a^{\text{max}}$, where a plastically active atom was detected. In order to refer the extension of the plastic zone to the position of the surface, we  add the indentation depth $d$ and thus identify the plastic-zone radius as $\Rpl=r_a^{\text{max}}+d$. This procedure is equivalent to that used for crystalline substrates, where the largest distance of a dislocation to the center of the contact area is used to determine the plastic-zone radius \cite{DBG05}. The plastic-zone size factor, $f$, \qeq{pl_factor} is displayed in Fig.~\ref{fig:f_factor} for maximum indentation and after removal of the indenter. The size factor assumes small values, around 1.2 at maximum indentation; these values are substantially smaller than for crystals where under similar indentation conditions, the minimum value of $f$ was found to be 2.6 \cite{ARU18}. 
Temperature influence is small; only above the glass transition, $f$ increases; here thermal activation processes initiated by the mechanical deformation continue and expand the initial damage deeper. As discussed above, the plastic zone systematically shrinks during indenter removal; here a stronger sensitivity to temperature is seen, 
as the more open structure generated during unloading  allows for  temperature-dependent structural relaxation.

Finally, we determine the influence of the indentation depth on the plastic density. We do this  at $T=0.1$ K, as
 at low temperatures the deformation mechanism is more anisotropic, see  Fig.~\ref{fig:Plastic}.   We note that at higher
 temperatures, the plastic-density decay is very similar for all indentation depths. 
 In Fig.~\ref{fig:Density-depth}(a), we superimpose the plastic zones formed at different indentation depths for $T=0.1$ K. Already at an indentation depth of 20~\AA, it is noticeable that the deformation is inhomogeneous and starts showing vein-like deformation below the indenter pointing at shear-band initiation. For deeper indentation, the veins do not grow  in  indentation direction; rather, new veins are  formed at the lateral side of the plastic zone.  
This means that a previously formed shear band  is not reactivated, in contrast to plastic processes during  uniaxial deformation of metallic glasses \cite{KML11}. 

The shear-band activity has an effect on the plastic-density distribution throughout the plastic zone as can be observed in Fig.~\ref{fig:Density-depth}(b) by the different way the plastic density  decays at different indentation depths. This effect can be better accounted for by the fit of Eq.~(\ref{comp_ex}) where the obtained values of $\beta$ can be seen to grow with indentation depth in the inset of Fig.~\ref{fig:Density-depth}(b).

Unsurprisingly, the thickness of the plastic zone, $r_a^{\text{max}}$, increases with increasing indentation depth. We plot the depth dependence of  the plastic-zone size factor $f$  in Fig.~\ref{fig:Density-depth}(c). We see that the increase of $f$ with depth is slow and can be described by a logarithmic depth dependence. Note that also size factors $<1$ appear, meaning that the extension of the damage in the direction of the indentation is smaller than in lateral direction.

We determine the plastic density in the core by averaging the values of the plateau from 10 \AA\ $<r_a<$ 15 \AA, see \qfig{fig:Density}b, and denote it as $\rho_{\text{max}}$. As \qfig{fig:Density}b shows, the core density slightly depends on temperature. Note that $\rho_{\max}$ in the core of  the plastic zone is almost as high as the atomic density $\rho_0$; this means that virtually every atom in the core region contributes to plastic rearrangement in the plastic zone.

Additionally, we correlate the core densities with the indentation hardness values determined for the same samples in Ref.\ \onlinecite{AKAU19}. These results are shown in the inset of Fig.~\ref{fig:f_factor}. There is a strong effect on the hardness even though the changes in the plastic density with temperature  are minimal. 

In summary, we studied the plastic zone created by nanoindentation of a CuZr  metallic glass using MD simulation. 
We defined plasticity by the von-Mises shear strain and calculated the density of  the plastically active atoms and the size of the plastic zone.
We found that the plastic-density distribution decays similarly at most temperatures. The plastic-zone size was found to be considerably  smaller than the one found under similar indentation conditions (indenter radius, indentation depth) in crystals. This  is a consequence of the mainly homogeneous deformation of the atomic regions around the indenter. Adjacent to the indenter a dense core zone exists where densification is slightly temperature dependent and strongly correlates with hardness.  

Deviations from an isotropic and homogeneous plastic zone occur in the form of shear bands. These are formed only at low temperatures and at sufficiently large indentation depths.  Also,  shear bands do not grow under continuing indentation but are swallowed in the plastic zone, and other shear bands are created instead.  
At low temperature, the plastic-density distribution decay shows a pronounced dependence on indentation depth. With increasing  indentation depth, the distribution decays faster, as  shear bands are activated. 


\begin{acknowledgments}

We acknowledge support by the Deutsche Forschungsgemeinschaft via the SFB/TRR 173. 
Access to the computational resources provided by the compute cluster `Elwetritsch' of the University of Kai\-sers\-lau\-tern is appreciated. 

\end{acknowledgments}

\bibliography{../../all,../../publ,../../Glasses}

\end{document}